\documentclass[utf8,english]{manuscript}
\usepackage{csquotes}
\usepackage{graphicx} 

\usepackage{amsmath} 

\usepackage{booktabs} 

\usepackage{longtable}

\usepackage{array}
\newcolumntype{L}[1]{>{\raggedright\let\newline\\\arraybackslash\hspace{0pt}}m{#1}}
\newcolumntype{C}[1]{>{\centering\let\newline\\\arraybackslash\hspace{0pt}}m{#1}}
\newcolumntype{R}[1]{>{\raggedleft\let\newline\\\arraybackslash\hspace{0pt}}m{#1}}

\usepackage{hyperref}
\hypersetup{
    bookmarksopen=true,     
    bookmarksnumbered=true, 
    linktocpage=true        
}

\begin{document}

\title{Unifying a Public Software Ecosystem: How Omaolo Responded to the COVID-19 Challenge}

\keywords{Public Software Ecosystem, Strategic Alliances, Crisis Response, Public Sector Innovation}

\abstract{%
  Public actors are often seen as slow, especially in renewing information systems, due to complex tendering and competition regulations, which delay decisions. This challenge is even greater in multi-company ecosystems. However, when faced with a common threat, the ecosystem needs to unite to face the challenge.
  
  This study explores how the Omaolo ecosystem in Finland evolved from traditional public-private cooperation to an alliance model during the COVID-19 pandemic from 2020 to 2022. It highlights how the crisis accelerated changes in operations and collaboration between public and private participants, identifying key shifts, benefits, and challenges.
  
  Key findings include the removal of traditional barriers and the creation of an alliance approach that sped up the development of Omaolo's symptom assessment tool. This improved collaboration, service scalability, and responsiveness to healthcare needs despite the initial regulatory and stakeholder alignment challenges.
  
  The study concludes that crises can drive agile responses in public ecosystems. The new collaboration model helped Omaolo to adapt quickly to changing service demands, managing healthcare patient loads more effectively. These findings highlight the value of flexible, collaborative strategies for responding to emergencies in public software ecosystems.
}

\author{Taija Kolehmainen\textsuperscript{1}}
\affil{\textsuperscript{1}University of Jyväskylä, Faculty of Information Technology, Jyväskylä, Finland}

\author{Reetta Ghezzi\textsuperscript{1}}

\author{Sami Hyrynsalmi\textsuperscript{2}}%
\affil{\textsuperscript{2}Lappeenranta-Lahti University of Technology LUT, Lahti, Finland}

\author{Tommi Mikkonen\textsuperscript{1}}

\author{Samuli Pekkola\textsuperscript{1}}

\author{Manu Setälä\textsuperscript{3}}
\affil{\textsuperscript{3}Solita, Tampere, Finland}

\type{This is the preprint version of the manuscript, awaiting formal peer review and publication.}
\maketitle


\chapter{Introduction}

Digital platforms, such as the smart grid, various digital services for healthcare, banking, or shopping, cloud platforms, and Internet of Things platforms, are intertwined in the business and citizens’ everyday lives. The success of businesses and public sector organizations depends on their ability to exploit new technologies and the social capacities afforded by the platforms. 

Platforms are often developed by a network of organizations, each contributing with their own services and components. Together, these components aggregate the platform. When the platform is established, the organizations agree on the means and methods, for instance, how the platform should be constituted, what services are included, and, importantly, how the development activities will occur. Public platforms designed and developed for digital public services are constrained and driven by rules, regulations, and explicit agreements.

This is an adequate approach in a relatively static situation. 

The global COVID-19 pandemic changed the game. It thrust the capabilities and resilience of healthcare systems, public services, and digital platforms into the spotlight. Previous agreement-based approaches with numerous, time-consuming quality controls and rigid agreements defining goals, methods, and schedules became inappropriate overnight when people's lives were put on a plate. 

This article tells the story of the evolution of the software ecosystem of \textit{Omaolo}, an e-health service where Finnish citizens could, among other contents, check whether their symptoms were severe and would require medical assistance or hospitalization. The service development began in 2016 when the Finnish government granted funding to fourteen municipalities or joint municipal authorities, a medical content provider,  and two IT companies to implement some basic features of the healthcare platform. The development proceeded slowly, each partner constantly securing their own backs, ensuring correct diagnoses, and producing high-quality services. An external shock, COVID-19, changed the situation, and the common threat significantly sped up development. The old approach to platform development with strict rules, practices, and divisions of labor was replaced by \textit{an alliance model} where either everybody wins – or everybody loses. The alliance model refers to the flexibility and adaptability of the ecosystem and its common goals, shared visions, and risks. Importantly, the model rewards the participants for achieving the objectives with bonuses or penalizes them for exceeding the budget or schedule or failing the features. 

External incidents or shocks, such as COVID-19 or other changes in the organization's operational environment, and internal incidents, such as acquisitions and mergers or organizational changes, are not uncommon. Under the circumstances, the organizations need to react to them somehow. For example, Smolander et al. \autocite{smolander2021heroes} identified four modes of collaboration in large enterprise systems development, each prevailing under different conditions and shifting to another when an incident occurs. These changes, however, necessitate organizational ability and agility. 

As a technical contribution, the article studies how the relationship between the organizations developing the Omaolo service evolved from a network-based approach to an approach resembling an alliance during the early months of the COVID-19 pandemic. Our overarching goal is to extract valuable insights that can inform future preparedness and resilience strategies, extending beyond technical considerations to encompass the actors' roles, incentives, regulations, business, and software within the ecosystem.

\section{BACKGROUND}

\subsection{Public Software Procurement}

Public sector organizations are bound by procurement policies when acquiring information and communication technology products and services, including software. The primary objective of public procurement guidelines is to enforce transparency, fairness, and cost-effectiveness throughout the procurement process. Guidelines, implied by the policies, typically encompass criteria for vendor selection, contract negotiations, and software management. The adoption of these policies may differ at the national level because there are no universally recognized international standards for software procurement.

In most countries, administration and government functions are divided among multiple agencies. Typically, each agency procures its own software solutions. This is done by issuing tenders for complete systems, often resulting in monolithic systems with limited interfaces for reuse. Maintenance tasks are typically tied to the selected vendor, who is responsible for maintaining the system throughout its operational life cycle.

An alternative way to develop such software has been proposed by \autocite{ghezzi2023towards}. The authors argue for improved resilience in public software systems when several vendors can participate in the development. This, in turn, means that a government-led ecosystem is formed to develop and maintain public sector software.

\subsection{Public Sector Digital Ecosystems and Their Governance}

Ecosystem thinking allows public procurement participants to pool resources to optimize the software's reusability and interoperability and enhance service delivery. In practice, this could mean complying with common standards and providing compatible and complementary solutions. It also encourages innovation and agile practices to enhance productivity, reduce costs, provide competitive edge solutions, and enable shared issue-solving.

Each element in the ecosystem influences and is influenced by the others, resulting in a complex network of inter-dependencies \autocite{peltoniemi2004business}. For instance, changes in one part of the ecosystem, such as introducing new software, can send ripples throughout the ecosystem, demanding adjustments in other elements. Maneuvering in these complex and interdependent settings, organizations need to move towards a more holistic and dynamic mindset instead of focusing on controlling their current resources. Generally, some of the requirements for creating value in ecosystems could be, for example, the following:
\begin{itemize}
    \item Enhanced integrability and standardization \autocite{sklyar2019organizing}.
    \item Open and adaptive resource integration \autocite{sklyar2019organizing}.
\item Establishing common goals and compatible incentives \autocite{adner2017ecosystem}.
\item Improving agility \autocite{dattee2018maneuvering} and multilateral compatibility \autocite{adner2017ecosystem}.
\item Fostering partnerships and flexibility in ecosystem management \autocite{iansiti2004strategy}.
\item Collaborative value creation, for example, through innovation \autocite{iansiti2004strategy}.
\item Finally, digitalizing services is essential for creating an ecosystem \autocite{sklyar2019organizing}.
\end{itemize}

Creating a collaborative and interconnected ecosystem requires organizations to commit to shared objectives and decision-making and to have appropriate structures, rules, and practices. Ecosystem governance refers to the development, management, and control of shared processes, operating models, practices, principles, and rules that enable the formation of such ecosystems \autocite{laatikainen_li_abrahamsson_2021}. In the public settings, decision-making and governance tend to be centralized and government-led. However, to maintain agility and enable innovation, some decentralization should be allowed within the agreed limits. 

In this study, we follow how a networked governance model, where each participant contributed from their own perspective, evolved into an alliance model approach. The alliance model is a project delivery method used in public procurement in Finland, where the governmental agencies closely partner with private companies \autocite{pekkala2022hankintojen}. In the model, the alliance acts as a cohesive team, under the terms agreed in the contract, to complete the project so that the jointly set and agreed objectives are met \autocite{jefferies2014using}. Often, the alliance model is "no blame, no disputes,'' meaning that the parties must be able to trust and support each other \autocite{jefferies2014using}.  In the alliance model, part of the cost risk is in the implementing company, i.e., an alliance is an agreement between two or more parties who take on a project jointly and severally, with shared profit and loss. The incentive scheme ensures that everyone works well together and focuses on making the project successful  \autocite{jefferies2014using}. The COVID-19 pandemic accelerated the formation of new strategic alliances across different sectors to address the immediate demands of, e.g., core healthcare
\autocite{cojoianu2020strategic}. Here, the role of governments is key in guiding the overall vision for both the immediate and longer-term needs \autocite{cojoianu2020strategic}.

\subsection{Omaolo and Its Ecosystem}

Omaolo is an electronic service and interaction channel for social and health care that supports self-care and self-help and directs individuals to appropriate assistance. The starting point was the government's objective to increase self-service and automation in the social and healthcare service model in 2016. With Omaolo, citizens can easily assess the type of care needed and receive personalized guidance or, if necessary, send contact requests. The system includes symptom assessment, health check-ups, and comprehensive well-being coaching to promote overall health. Social service assessments help determine eligibility for specific services.

Omaolo is a CE-marked medical device from May 2022 onwards, which complies with the EU Regulation 2017/745 (MDR) requirements \autocite{mdr}. The CE-marking signifies exceptional quality and safety and of their documentation for citizens using Omaolo and health and social care professionals. Consequently, the Omaolo system can provide up-to-date information to improve the effectiveness of social and health services and build different service channels, thus facilitating interaction between citizens and health professionals.

Omaolo involves several stakeholders having different roles, responsibilities and incentives in the ecosystem. These are briefly summarized in Table~\ref{tab:ecosystem-participants}.

\begin{longtable}{|p{.16\textwidth}|p{.17\textwidth}|p{.28\textwidth}|p{.28\textwidth}|}
    \hline
    \textbf{Actor} & \textbf{Role} & \textbf{Responsibility} & \textbf{Incentives}\\
    \hline
    \hline
    Finnish Government & Financing Body; Policy Maker & Ensure healthcare services are available to citizens. Provide funding and strategic direction aligned with national healthcare priorities. & Promote enhanced utilization of e-services, encompassing self-care and counseling, to support citizen engagement. \\
    \hline
    In-House Company & Ecosystem Coordinator and Facilitator & Provide and administrate public sector digital services for Finland, including Omaolo. Oversee project management and coordination between stakeholders and foster collaboration. & Support digitalization of healthcare services. Enhance public sector efficiency. \\
    \hline
    Medical Content Expert Organization & Content and Knowledge Provider & Provide medical content knowledge base for the service. Continually update medical content for accuracy and relevance. & Contribute to public health knowledge and leverage expertise in digital healthcare solutions. \\
    \hline
    Companies & Vendors; Technical Solution Providers & Provide a range of IT services to DigiFinland to improve existing offerings and implement new solutions to ensure innovation, agility, and responsiveness to healthcare needs. & Increased sales and increased portfolio via expanded market presence and technological innovation. \\
    \hline
    Regulatory and Standard-Setting Organizations & Healthcare Regulatory Authorities; Device Certifiers & Protect public safety through regulatory actions, ensuring compliance with healthcare regulations, and certifying medical devices. & Legal obligations. Maintain high standards of healthcare quality and safety in digital solutions. \\
    \hline
    Regional Operators (municipalities, well-being service counties) & Public Healthcare Service Providers and Organizers & Operationalize services to citizens. Implement and adapt services to local healthcare needs, engaging with communities. & Optimize the allocation of healthcare professional resources for greater efficiency. Improve local health outcomes by delivering services that meet regional demands. \\
    \hline
    National Health Agencies & Healthcare Strategy, Policy Advisors, and Regulatory Contributors & Guide health policies, advise on strategy, and contribute to regulatory processes. Evaluate the public health impact of Omaolo. & Effective healthcare policies and successful implementation of health strategies. \\
    \hline
    Healthcare Professionals & Service Users & Participate in service co-development wherever possible and provide clinical feedback for improvement. & Enhance practice efficiency and patient care. \\
    \hline
    Citizens & Service End-users & Use health care services responsibly by choosing a suitable channel for engaging. Provide feedback for improvement. & Decide when to resort to self-care and when to contact healthcare professionals based on convenience, effectiveness, and responsiveness of the services. \\
    \hline%
    \caption{Key Omaolo ecosystem actors, their roles, responsibilities, and incentives.}
    \label{tab:ecosystem-participants}
\end{longtable}

\section{RESEARCH APPROACH}

This study examines the Omaolo service as a public software ecosystem. The study is thus a case study of the Omaolo platform and its management during the COVID-19 pandemic from 2020–2022. We examine, particularly, the changes that took place in the administrative roles and responsibilities of the platform as a result of the pandemic-caused crisis.

To understand the ecosystem evolution, we conducted a set of nine semi-structured interviews and gathered publicly available documentation. The interviews were carried out during Fall 2023, by which time the pandemic was considered as no longer an endangering disease in Finland. Table \ref{tab:interviews} lists our informants.

Deriving the interviewees' experiences, we modeled the Omaolo ecosystem using the Ecosystem Governance Compass, a domain-specific modeling language \autocite{sroor2022modeling} enabling visual modeling of ecosystem components, interactions, and dependencies between ecosystem participants. The Ecosystem Governance Compass maps the building blocks of a digital business ecosystem from governance, business, and technology perspectives. It examines the regulation of cooperation and the obligations of actors in a legal context (following Laatikainen et al. \autocite{laatikainen_li_abrahamsson_2021}). The resulting models support analyzing and exploring a complex ecosystem via knowledge integration \autocite{Mader_Wupper_Boon_Marincic_2008}.

We particularly focused on understanding the ecosystem governance structures and model, interactions and inter-dependencies between the actors, responsibility distribution among the roles, and decision-making processes. By modeling relevant components and their causal and dependent interactions, we were able to identify the ecosystem structures and governance mechanisms  \autocite{Mader_Wupper_Boon_2007}.
\clearpage
\begin{longtable}{|p{0.07\textwidth}|p{.15\textwidth}|p{.2\textwidth}|p{.05\textwidth}|p{.39\textwidth}|}
    \hline
    \textbf{ID} & \textbf{Organization} & \textbf{Particant Role} & \textbf{Min} & \textbf{Key Topics}\\
    \hline
    \hline
    INT01&Vendor&Project Lead, Designer&63&Governance, COVID-19 response, Values, Cooperation, Technology, Innovation, Regulations\\\hline
    INT02&Public Sector&Medical Director&54&Governance, COVID-19 response, Values, Cooperation, Technology, Innovation, Regulations\\\hline
    INT03&Public Sector&Medical Director&54&Governance, COVID-19 response\\\hline
    INT04&Public Sector&Portfolio Manager&82&Governance, COVID-19 response, Values, Cooperation, Technology, Innovation, Regulations\\\hline
    INT05&Vendor&Senior Software Engineer&74&Values, Cooperation, Technology\\\hline
    INT06&Public Sector&Head of Operations&36&Business, COVID-19 response\\\hline
    INT07&Public Sector&Senior Software Engineer&61&Governance, COVID-19 response, Values, Cooperation, Technology, Innovation, Regulations\\\hline
    INT08&Non-Governmental Organization&Development Manager&49&COVID-19 response\\\hline
    INT09&Non-Governmental Organization&Medical Director&88&Governance, COVID-19 response, Values, Cooperation, Technology, Innovation, Regulations\\
    \hline%
    \caption{Interview subject, their background, recording duration and key topics discussed.}
    \label{tab:interviews}
\end{longtable}

\section{FINDINGS}

\begin{figure}[ht]\centering
  \includegraphics[width=1\linewidth]{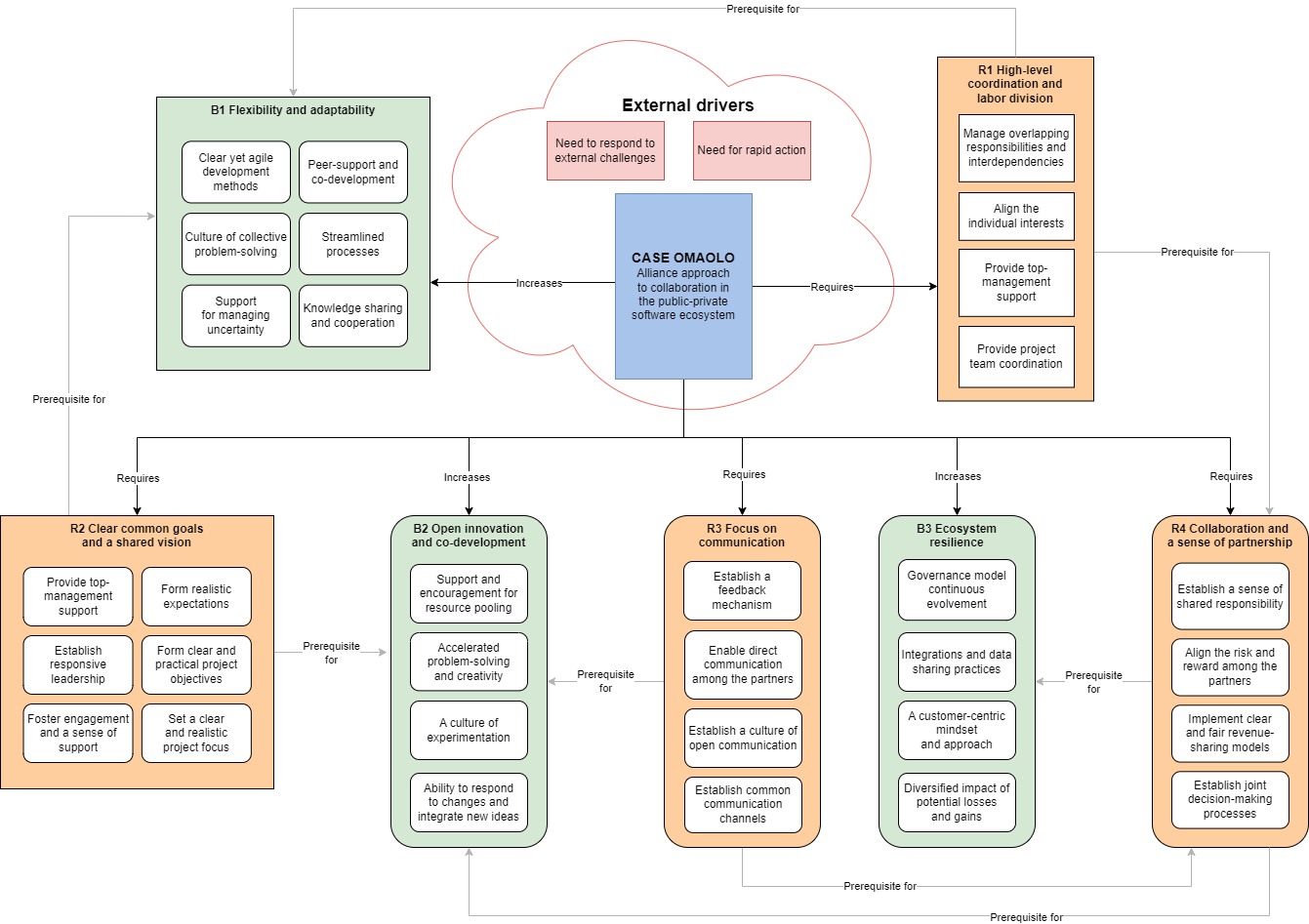}
  \caption[Benefits (B) and requirements (R) for alliance approach in the case Omaolo
  \parencite{fig:results-benefits-requirements}]{Benefits (B) and requirements (R) for alliance approach in the case Omaolo.}
  \label{fig:results-benefits-requirements}
\end{figure}

An external incident or threat, the COVID-19 pandemic, forced the Omaolo ecosystem to evolve from traditional public-private cooperation towards an alliance approach. When the pandemic hit, the development focus was quickly directed toward the Omaolo symptom assessment. The sense of urgency and purpose shaped the ways in which the ecosystem participants addressed the common threat and took action in a rapidly changing situation. They had a common goal of sustaining Finland's healthcare system even when the patient load skyrocketed. The participants removed the boundaries between public and private organizations, started to work very closely together, and aligned their service development efforts. In this paper, we identified three benefits (B1–3, see Figure \ref{fig:results-benefits-requirements}) that emerged from the change of focus and collaborating effectively. Also, four strategic requirements (R1–4, see Figure \ref{fig:results-benefits-requirements}) must be in place for such an all-win-or-all-lose alliance to play out. These changes were not without challenges, as the ecosystem had to manage evolving situations and regulatory complexities while developing regulated medical software.

First, flexibility and adaptability (B1) were emphasized when navigating rapidly changing conditions. The urgency of the COVID-19 pandemic resulted in streamlining the processes, particularly in tendering and agreement protocols, and led to evident behavioral changes among the contracting entities and policymakers. For example, direct negotiations replaced the usual tendering and bidding processes, and flexibility was embraced over strict adherence to standard procedures. Public sector entities strategically utilized existing framework contracts with vendors instead of issuing new tenders. Policymakers and healthcare providers displayed adaptability by enabling a more pragmatic approach, deviating from conventional 'by the book' methods to meet the project's urgent demands. To ensure fast software delivery, there was a need to coordinate and align conflicting interests (R1), for example, in meeting the medical device regulation requirements. Developing regulated medical software at a rapid pace favored those companies that had strong regulatory know-how and predefined processes for managing regulatory compliance. Further, shared goals and a vision (R2) between the vendors and the coordinator facilitated the determined deployment of the Omaolo e-health service. The common alignment enabled bold actions and agile adjustments in the service based on user feedback. Finally, efficient data sharing among vendors, healthcare professionals, and government entities ensured that the service was in line with Finland's evolving national COVID-19 strategy. Coordination efforts included, for example, the medical director maintaining the dialogue with various medical actors throughout the project (R1). However, despite facilitating information exchange between the stakeholders, some change requests were received simultaneously with the public. This then required immediate updates to the software.

Second, the need for rapid solution development and deployment promoted open innovation and co-development (B2). The actors sought ways to collaborate and address changing demands. This necessitated immediate and close collaboration, which was perceived as effective and beneficial by all parties. Meetings were kept short and informal, and anyone with relevant knowledge could join at a low threshold (R3). During the most critical times, the actors concentrated entirely on updating the syndrome assessment content, shared the same objectives (R2), and postponed the development of other Omaolo features. Omaolo's highly automated assessment of symptoms and laboratory test scheduling in the most affected regions significantly influenced the implementation of the national COVID-19 strategy and the effective management of patient loads. Further, this shift in collaboration (R4) facilitated changes in the ways of working, for example, supporting remote work and flexible office hours. The sense of shared responsibility also resulted in ignoring previous disagreements and putting them aside. In the interviews, each vendor reported working long hours, prolonging into evenings and weekends, and demanding commitment from both teams and their families (R4).

Third, the pandemic required quick adjustments, especially in coordination (R1) and communication (R3), to increase ecosystem resilience (B3). The crises brought focus to hierarchical and organizational structure in the ecosystem, highlighting the importance of agile and joint decision-making (R4). Initially, the COVID-19 response lacked clear specifications, and decisions had to be made on the fly to meet the urgent demands set for the Omaolo service. The responsibility of balancing between the quality and speed of deployment was on each individual actor. Later, the collaboration practices became more structured, with clear roles, responsibilities, and schedules for the development. Further, there had been earlier attempts to centralize communication in the Omaolo ecosystem before the pandemic. However, such centralization in communications quickly became ineffective as the ecosystem faced a common threat, resulting in alternative channels and secretive communication for technical issues. This led to a rapid change in communication strategies when faced with the common threat, emphasizing the need for flexibility in crisis management. Communication became more direct and effective (R3), connecting those who understood the issues with those who could resolve them. Another success factor was the readily available and quickly formed deployment pathways and processes for the symptom assessment in two hospital districts, as well as a well-functioning network for information sharing. From the healthcare point of view, the service had concordant practices and care paths. However, while health professionals were only partially familiar with the Omaolo pilot project before the pandemic, introducing new functions was straightforward in the most affected areas due to, e.g., active promotion of Omaolo and prioritizing digital services.

The race against the pandemic caused several challenges. One of the main difficult points was adjusting Omaolo with the national strategies, which were changing fast and requiring frequent updates. This urgency created high-pressure work conditions with extended working hours, impacting not only the development teams but also their personal environments. Balancing the immediate focus on the COVID-19 response – while simultaneously resuming the broader development of the Omaolo e-health services –  demonstrated the challenges of managing a critical healthcare project under quickly changing circumstances. In addition, the COVID-19 project navigated through regulatory complexities with mixed opinions and interpretations and under heavy time pressure. Some found it burdensome and costly, while others appreciated the structure it provided, standardizing and assuring software quality. However, limited resources slowed the Omaolo development process and impacted the software quality, as regular updates were essential. The service was made freely available to all user organizations and centrally financed by the government to ensure equal service throughout Finland when meeting the rapid demands of the COVID-19 emergency. During the pandemic, the short-term profitability of the service was not a concern. 

In summary, the Omaolo case illustrates the potential of the alliance model in terms of flexibility and adaptability, ecosystem resilience, and supporting innovation and faster responses in public healthcare software development. Our findings indicate that the emphasis on common goals and shared vision, communication, collaboration, and coordination significantly contributed to the successful development of the Omaolo software.

\section{DISCUSSION}

The COVID-19 pandemic changed society and people's ways of working, as well as how the digital ecosystem operated. With the Omaolo case, an alliance approach to mobilize a public-private partnership transformation and enable the development of COVID-19-specific features to the Omaolo e-health service emerged during the pandemic in the years 2020–2021. Although the structure and management practices of the Omaolo ecosystem have evolved throughout its existence, the pandemic initiated fundamental changes. These changes impacted the progress and stability of the system during the pandemic. 

The platform ecosystem's ability to react to the pandemic required organizational resilience from all parties. Such responsiveness and change are not easy and evident. Abandoning old rules, regulations, work practices, communication means, individual incentives, business interests, and traditions and replacing them with a shared incentive and straightforward get-things-done-quickly attitude and practices did not happen easily. With Omaolo, the pandemic drove the parties, and frankly speaking the whole country, into a situation where the healthcare system collapses if all citizens enter the hospitals en masses to check whether they have severe COVID-19. Under these circumstances, the need for self-service through Omaolo became vital. The pandemic was an external incident, a shock, or a catalyst (compare to Smolander et al. \autocite{smolander2021heroes}) that put the ecosystem in a position to be a hero and save the country – or fail. This intrinsic motive in the Omaolo ecosystem allowed the parties to cut corners in every possible way.

In addition to external shock and internal motivation, the change also entails adequate cultural, educational, and societal background and context. In the spirit of Christensen and Eyring~\autocite{christensen2011innovative}, organizations, and their employees cannot respond to changes quickly unless such ability is built in their DNA. Luckily, the ecosystem DNA allowed this.
The alliance model postulates open information and transparency in the relationship between public and private partners \autocite{jefferies2014using}. This was also experienced in the Omaolo case. To be flexible with rules, regulations, collaboration and work practices, and individual working hours, the parties had to, for example, bend the competition rules and individual profitability goals – and to be completely honest and open with them. The overall transparency increased in the ecosystem. Whether this transparency and practices are sustainable remains to be seen. The need for including the public sector actors is, however, emphasized in the literature \autocite{cojoianu2020strategic} and in our case.

\section{CONCLUSION}

In conclusion, adopting an alliance model in software ecosystems can significantly enhance the ecosystems' ability to respond to an emergency by emphasizing open communication, shared goals, and transparency. Aligning stakeholder resources and expertise and simplifying decision-making processes facilitates joint development and rapid innovation between stakeholders. Furthermore, encouraging flexibility in organizational rules, work practices, and regulations can promote resilience and adaptability in crisis situations. Streamlining processes enables rapid response to changing demands and ensures the deployment of essential solutions.
The common threat and a sense of urgency created genuinely new ways of working together to deliver welfare and healthcare software. These included maintaining communication and information sharing, combining agile development approaches with regulated medical software, streamlining processes, direct state funding, and existing deployment paths.

In doing so, enhancing transparency improves collaboration efficiency and trust among all stakeholders in digital ecosystems. The open sharing of information, progress, and challenges across all levels of partnerships, especially in crisis response, is crucial to enable rapid adaptation of solutions, pre-empt potential conflicts, and foster a culture of mutual accountability and collaboration.

\printbibliography


\end{document}